\documentclass[aps,prl,groupedaddress]{revtex4}

\usepackage{keyval} \usepackage{dcolumn}

\usepackage[]{amsmath}

\def\epl{Europhys.\ Lett. }

\def\prl{Phys.\ Rev.\ Lett.}
\def\prb{Phys.\ Rev.\ B }
\def\pr{Phys.\ Rev. }

\usepackage[psamsfonts]{amsfonts}
\usepackage[psamsfonts]{amssymb}
	
\usepackage{graphicx}

\usepackage{bm}

\usepackage[usenames,dvipsnames]{color}

\usepackage[normalem]{ulem}

\newcommand{\threej}[6]{\ensuremath{\left(\begin{array}{ccc}#1 &#2 &#3 \\ #4 &#5 &#6 \end{array}\right)}}
\newcommand{\sixj}[6]{\ensuremath{\left\{\begin{array}{ccc}#1 &#2 &#3 \\ #4 &#5 &#6 \end{array}\right\} }}

\begin{document}

\title{Exchange energy dominated by large orbital spin-currents in $\delta$-Pu}

\author{Francesco Cricchio}
\affiliation{Department of Physics and Materials Science, Uppsala University, Box 530, SE-75121 Uppsala, Sweden}
\author{Fredrik Bultmark}
\affiliation{Department of Physics and Materials Science, Uppsala University, Box 530, SE-75121 Uppsala, Sweden}
  \author{Lars Nordstr\"om}
\affiliation{Department of Physics and Materials Science, Uppsala University, Box 530, SE-75121 Uppsala, Sweden}

\date{\today}

\begin{abstract}
The electronic structure of the anomalous $\delta$-phase of Pu is analyzed by 
a general and exact reformulation of the exchange energy of the $f$-shell. It is found that the dominating contribution to the exchange energy is a polarization of orbital spin-currents that preserves the time reversal symmetry, hence a non-magnetic solution in accordance with experiments. The analysis brings a unifying picture of the role of exchange in the $5f$-shell with its relatively strong spin-orbit coupling. The results are in good accordance with recent measurements of the branching ratio for the $d$ to $f$ transition in the actinides.
 \end{abstract}

\maketitle

The electronic structure of the actinides stands out among the elements of the periodic table as most intricate. Their 5$f$ states form narrow bands with a comparatively large spin-orbit coupling. Rather subtle changes in the electronic structure leads to different ground states, and in various actinide compounds the 5$f$ show itinerant, localised as well as heavy-fermion behaviour and are responsible for both magnetic order and superconductivity. In recent years this complexity has been exemplified by various attempts to theoretically understand the phase diagram of Pu, and especially the formation of its high temperature, large volume, highly anomalous $\delta$-phase \cite{Soderlind1,Savrasov1,Shick,Anisimov,Savrasov2,Kotliar,Eriksson,Soderlind2}.
As a first progress, it was observed that the stability of this phase can be understood if allowed for spin polarisation (SP) \cite{Soderlind1,Savrasov1}. However, the existence of any magnetic moment are in contrast to a large amount of experimental observations \cite{expts}. Most recently it was pointed out that orbital polarisation (OP) plays a major role in stabilising this magnetic solution \cite{Soderlind2}. Then, by utilizing the so-called LDA+U approach,  i.e.\ a local density approximation to density functional theory with an added extra local Hartree-Fock (HF) term. When using a HF term in its most general form allowing for off-diagonal spin coupling \cite{Shick,Anisimov}, it was observed that a non-magnetic large volume phase could be stabilised in a somewhat counter intuitive way -- by increasing the exchange energy the moments  vanish. In the analysis of these calculations, it was pointed out that the solution was closer to a {\em jj} coupling scheme than the usual {\em LS} coupling and that the configuration was closer to a $f^6$ than the expected $f^5$. An observation of such a  configuration is again opposing many experimental findings, although some recent experiments verify a {\em jj}-like coupling \cite{vdLaan1,Moore}. 
Meanwhile there have also been various dynamical mean field theoretical (DMFT) calculations, which also lead to a high volume non-magnetic state, at least when allowing for the off-diagonal spin coupling, which they attribute to a localization of the 5$f$ states. There are also a model of mixed valency, where part of the 5$f$ states are localised while the rest are itinerant \cite{Eriksson}.  With four localised and one itinerant a large volume phase is calculated to be stabilized with an electronic structure close to what is observed by photo-emission experiments \cite{PES}.

The purpose of the present Letter is to explore the relationship of the SP+OP approach with that of the general LDA+U approach with its more flexible exchange interaction, and to explain how the non-magnetic state is stabilised by the latter. This is accomplished be re-expressing the general exchange interaction of the LDA+U approach as a sum of interacting multi-poles. With the aid of this expression 
it is argued that in the case of $\delta$-Pu, the SP is overtaken by a variant of OP that does not break the time reversal symmetry as is confirmed by electronic structure calculations within the LDA+U approach.
In this context, there is a discussion of the nature of this exact expression for the OP and how it compares with earlier formulations \cite{Brooks,Eriksson2}. Finally, the corresponding OP multi-pole is discussed in connection with existing $d$ to $f$ branching ratios experiments \cite{vdLaan1,Moore}, where it appears through a sum rule.

In the most general version of LDA+U \cite{solovyev} the HF correction enters with a Hartree (H) and exchange (X) term as
\begin{equation}
E_\mathrm{H}+E_\mathrm{X}=
\frac{1}{2}\sum_{abcd} \left(\rho_{ac}\rho_{bd}-\rho_{ad}\rho_{bc}\right)\,\left\langle ab|g|cd\right\rangle , \label{ldau}
\end{equation}
where $\rho_{ab}$ is one element of the density matrix for the $\ell^\mathrm{th}$ shell, with dimension$(4\ell+2)\times(4\ell+2)$, (or $2[\ell]\times2[\ell]$ if we use the conventional notation, $[\ell]=2\ell+1$) which acts as an occupation matrix. Here $a$ is a combined label for the magnetic quantum number $m_{a}$ and the spin variable $s_{a}$. The interaction has the form
\begin{align}
\left\langle ab|g|cd\right\rangle =& \delta(s_{a},s_{c})\delta(s_{b},s_{d}) [\ell]^2 \nonumber\\
\times& \sum_{kq} (-)^q \,F^{(k)} c^{(k)}(m_{a},m_{c}) c^{(k)}(m_{b},m_{d}).
\end{align}
where
$c^{(k)}$ are Gaunt coefficients
and $F^{(k)}$ are the Slater integrals of the screened Coulomb interaction, in this work we stay with the convention and choose the latter on physical grounds. In the exchange term the spin Kronecker-deltas will allow for a non-diagonal spin interaction between the two density matrices $\rho$, giving rise to a spin-mixing. 

\begin{figure}[htbp]
\begin{center}
\includegraphics[width=1\columnwidth]{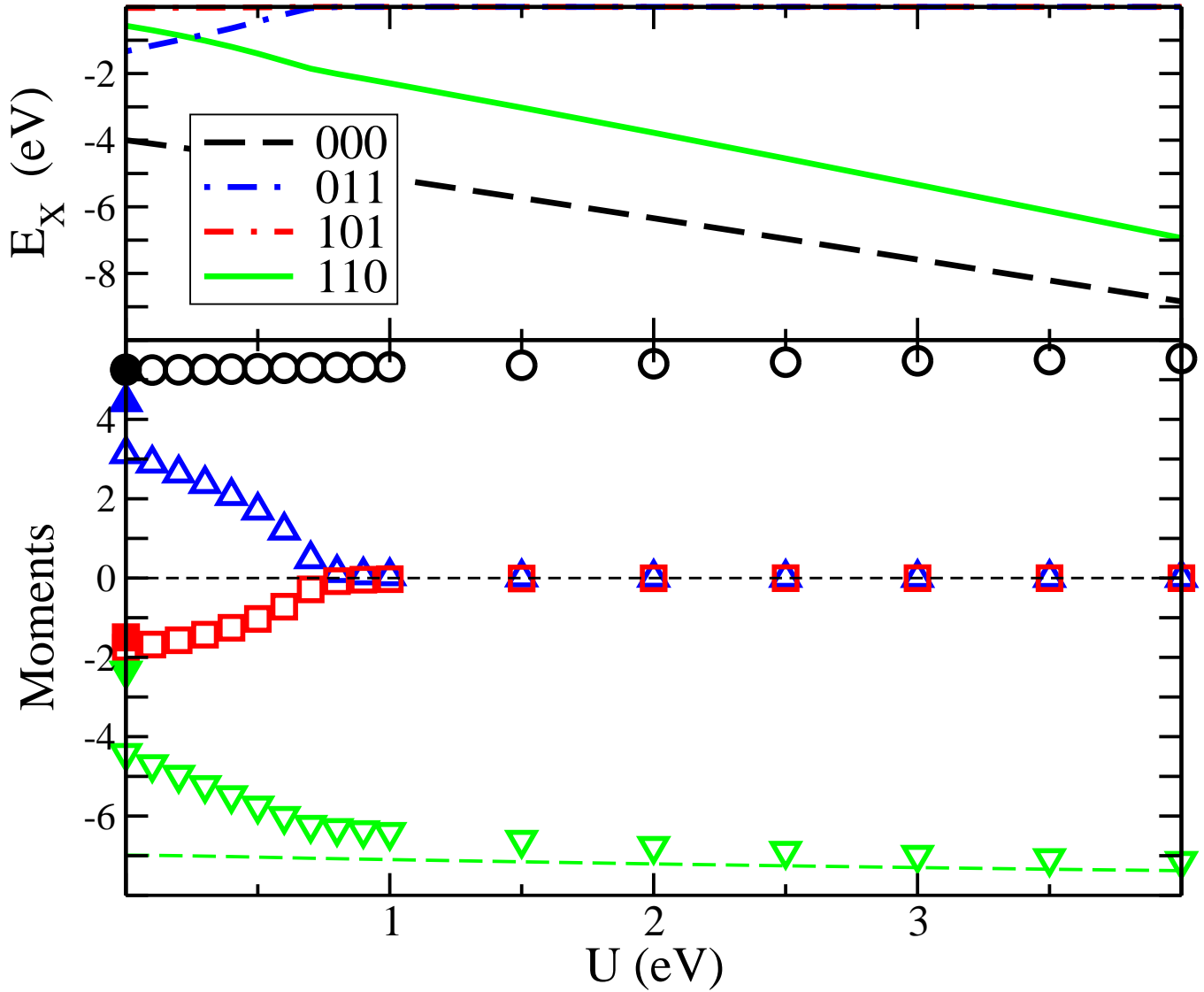}
\caption{Moments and exchange energies from LDA+U-AMF calculations of fcc Pu ($a$=4.64\AA) within the FP-APW method. Basis set cutoffs corresponding to $R_\mathrm{MT}G_\mathrm{max}=9$, with muffin tin radius $R_\mathrm{MT}=1.56$\AA, and a Brillouin zone sampling of 864 points for a two atom cell were used. The Slater parameters are summarized in two parameters $U$ and $J$ in the same way as in Ref.\ \cite{Shick}. The spin (blue triangles) and orbital (red squares) moments are shown for a varying $U$ but a constant $J=0.68$ eV, which corresponds to $E^{(3)}=53$ meV, in the bottom part. Filled symbols indicate the LDA results. Also displayed are the $5f$ occupation number (black) as well as the $\mathbf{w}^{110}$ (green) where the dashed green line indicates the corresponding saturation limit.  At the top, the most significant terms of the exchange energy in Eq.\ (\ref{X-mom}) are displayed as lines, with same color scheme as for the moments,  with the tensor rank identifying each, given in the legend.}
\label{fig1:Pu}
\end{center}
\end{figure}

This method has been implemented \cite{Cricchio} in the full potential augmented plane wave (FP-APW) package {\sc Exciting} \cite{exciting}, and the results above have been verified. A straight-forward density functional approach leads to an anti-ferromagnetic order with large spin and orbital moments, while switching on the LDA+U HF interaction, including spin-mixing terms,  leads to a non-magnetic solution 
%
as displayed in Fig.\ (\ref{fig1:Pu}) for a double counting of the type around mean field (AMF) \cite{Shick}.

To straight-forwardly analyze this non-magnetic solution is cumbersome due to the $4[\ell]^2$ independent elements of the density matrix $\rho$. Instead we take the following approach.
As a generalization of the fact that one gets the spin and orbital moment from the density matrix as  e.g.\ $\langle{S}\rangle=\mathrm{Tr}\,{S}\,\rho$, we introduce the expectation value of double tensor operators  $\mathbf{w}^{kp}=\mathrm{Tr}\, \mathbf{v}^k\, \mathbf{t}^p\,\rho$ \cite{Racah-II,vanderLaan,comment}. In our basis the tensor operators take the form
\begin{align}
v^k_{x}&\equiv\left\langle m_{b}\right|v^{k}_{x}\left|m_{a}\right\rangle = (-)^{\ell-m_{b}}
 \threej{\ell}{k}{\ell}{-{m}_{b}}{x}{m_{a}} 
n_{\ell k}^{-1} \nonumber\\
t^p_{y} &\equiv\left\langle s_{b}\right|t^{p}_{y}\left|s_{a}\right\rangle=  (-)^{s-s_{b}}\threej{s}{p}{s}{-{s}_{b}}{y}{s_{a}}
n_{sp}^{-1}\,.
\end{align}
Here we have use the so-called 3$j$-symbols $\left(\dots\right)$ \cite{Judd}
 and the same normalisation as in Ref.\ \cite{vanderLaan}.


An attractive property of the multi-pole double tensors $\mathbf w^{kp}$ 
is their simple interpretations. As have been pointed out \cite{vanderLaan} the
$\mathbf w^{k0}$, with $k$ even, are related through Wigner-Eckhart theorem to the $k$-th multi-pole moment of the $\ell$ charge density 
while the $\mathbf w^{k1}$, with even $k$, are related to the multi-poles of the magnetization density.
Finally the odd $k$ correspond to currents, i.e.\ the tensors are related to the multi-pole moments of the current ($p=0$) and the spin-current ($p=1$).

It is fruitful to view the introduction of $\mathbf{w}^{kp}$ as a transformation of the density matrix $\rho$ to these double tensors. 
This transformation 
is one-to-one \cite{Lie}, which is consistent with the fact that the number of parameters are kept.
The sum of the rank of the tensor operators are $\sum_{k}[k]\times\sum_{p} [p]=4[\ell]^2$, which is equal to the number of independent {(real and imaginary)} components of $\rho$, since it is hermitian. So we keep the same information as in the density matrix, but now distributed on $[k][p]$ independent double tensors, which turns out to be very convenient.
The inverse transformation is readily obtained by
utilizing orthogonality properties of the 3$j$-symbols \cite{Judd},
\begin{align}
\rho_{ac}=&\sum_{kx}[k]n_{lk}(-)^{m_{c}-\ell}\threej{\ell}{k}{\ell}{-{m}_{c}}{x}{m_{a}}\nonumber\\
\times&
\sum_{py}[p]n_{sp}(-)^{s_{c}-s}\threej{s}{p}{s}{-{s}_{c}}{y}{s_{a}}w^{kp}_{xy}\label{tensor-to-dm}\, .
\end{align}

As can be straight-forwardly shown,
with the introduction of the double tensors $\mathbf{w}^{kp}$ and the relation Eq.\ (\ref{tensor-to-dm}), both the direct and the exchange term can be put in simple forms,
 \begin{align}
 E_{\mathrm H}&=\sum_{k}  F^{(k)} I(\ell,k,k)\,
  \mathbf{w}^{k0}\cdot \mathbf{w}^{k0}\label{H-mom}\\
E_{\mathrm X}&=-\sum_{k}F^{(k)} \sum_{k_{1}=0}^{2\ell}J(\ell,k,k_{1})   \sum_{p=0}^1 \mathbf{w}^{k_{1}p}\cdot\mathbf{w}^{k_{1}p}\label{X-mom}\, ,
\end{align}
with 
\begin{align}
I(\ell,k,k_{1})= \frac{[\ell]^2n_{\ell k_{1}}^2}{2} \threej{\ell}{k}{\ell}{0}{0}{0}^2
\end{align}
and
\begin{align}
J(\ell,k,k_{1})&=
I(\ell,k,k_{1}) \frac{(-)^{k_{1}} [k_{1}]}{2}  \sixj{\ell}{\ell}{k_{1}}{\ell}{\ell}{k} \label{vxv-param}
\, ,
\end{align}
where the $\left\{\dots\right\}$ symbol is the $6j$-symbol \cite{Judd}.
This is a remarkably simple reformulation of the exchange energy. It is a straight-forward generalisation of the Stoner-like exchange formulation of SP, with scalar products of quantities with themselves.
The final expression is actually almost identical to an expression derived by Racah for a $\ell^2$ configuration \cite{Racah-II}, but is more general in that it is valid for any non-integer occupation of the $\ell$-shell.
The calculated interaction strengths from Eq.~(\ref{vxv-param}) become even simpler when going from Slater parameters to a certain linear combination of them called Racah parameters, as many values then vanish. The relation between the two types of parameters is given for the case of $f$-shell in Table \ref{Racah} and the corresponding transformed interaction strengths are tabulated in Table \ref{parameters}.

\begin{table}
  \centering 
  \caption{The expansion coefficients $c(k,k')$ of the Racah parameters for the $f$-shell,
  $E^{(k)}=c(k,k')F^{(k')}$, are given. We can note the relations with the standard $U$ and $J$ parameters:
  $E^{(0)}=U-J$ and $E^{(1)}=\frac{7}{9}J$.}\label{Racah}
 %
  \begin{tabular}{|c|cccc|}
\hline
       & $k'$ & & &  \\
   $k$ & 0 &2 &4 &6  \\
 \hline
   0 & 1 & --{2}:{45} & --{1}:{33}&--{50}:{1287}\\
      1 & 0 & {14}:{405}&{7}:{297}&{350}:{11583} \\
      2 & 0 & {1}:{2025}&--{1}:{3267}&{175}:{1656369} \\
      3 & 0 &{1}:{135}&{2}:{1089}&--{175}:{42471}\\
     \hline
\end{tabular}
\end{table}

\begin{table}
  \centering 
  \caption{The interaction strenghts $\tilde{J}(3,k,k_{1})$ from the multi-pole expansion of the exchange energy in terms of Racah parameters, 
  $E^{(k)}$, 
 $E_{\mathrm{X}}=-\sum_{kk_{1}p} E^{(k)}  \tilde{J}(\ell,k,k_{1}) \mathbf{w}^{k_{1}p}\cdot\mathbf{w}^{k_{1}p}$. }\label{parameters}
\noindent\(
\begin{array}{|cccccccc|}
\hline
\hfill\vline& k_{1} &&&&&&\\
  \begin{array}{c}
  k
  \end{array}\hfill\vline&
0   & 1 & 2 & 3 & 4 & 5 & 6 \\
 \hline
 \begin{array}{c}
  \end{array}\hfill\vline&&&&&&&\\
 \begin{array}{c}
  0\\\\1\\\\2\\\\3
  \end{array} \hfill\vline& 
\begin{array}{c}
  \frac{1}{28} \\\\
  \frac{9}{28} \\\\
  0 \\\\
{0}
  \end{array}&
  \begin{array}{c}
   \frac{9}{112}\\\\
   0 \\\\
   0 \\\\
   \frac{297}{112}
   \end{array}&
    \begin{array}{c}
        \frac{25}{336} \\\\
    \frac{25}{168} \\\\
    \frac{3575}{168}\\\\
    -\frac{275}{336}
   \end{array}&
    \begin{array}{c}
     \frac{1}{24}\\\\
 0 \\\\
    0 \\\\
    0 
     \end{array}&
    \begin{array}{c}
     \frac{9}{616}\\\\
  \frac{9}{308}\\\\
   -\frac{585}{154}\\\\
   -\frac{9}{154}
    \end{array}&
    \begin{array}{c}
   \frac{1}{336}\\\\
   0 \\\\
   0 \\\\
   -\frac{3}{112}
    \end{array}&
    \begin{array}{c} 
        \frac{1}{3696}\\\\
   \frac{1}{1848} \\\\
   \frac{5}{264} \\\\
   \frac{1}{528}
    \end{array}\\
\begin{array}{c}
  \end{array}\hfill\vline&&&&&&&\\
\hline
\end{array}
\)
\end{table}

The independent terms in this exchange energy expansion have simple physical meanings, for instance the term with $\mathbf{w}^{01}$ is the Stoner-like SP, while the terms involving $\mathbf{w}^{k_{1}1}$ with $k_{1}=2,4,6$ are responsible for an intra-atomic non-collinear spin polarisation \cite{nordstrom}.
In particular, we have an explicit expression for orbital polarization, or Hund's second rule, for a general non-integer system.  
From Eq.\ (\ref{X-mom}) and with the observation that orbital moments are $\left\langle L\right\rangle
=\ell\mathbf{w}^{10}$,
we get a generalized OP term for $f$ states 
as
\begin{align}
E^{\mathrm{OP}}_{3} = -\frac{E^{(0)} +33 E^{(3)}}{112}\,\sum_{p}
9\,\mathbf{w}^{1p}\cdot\mathbf{w}^{1p}\, .
\end{align}
This expression is very close in resemblance  to the expression suggested by Brooks \cite{Brooks,Eriksson2} (OP-B), which in our terminology looks like 
\begin{align} 
E^\mathrm{OP-B}_{3} = -\frac{E^{(3)}}{4}\,\sum_{p}
\left(3\,{{w}^{1p}_{00}}\right)^2\,.
\end{align}
We note that there are three corrections to this simplified OP-B formulation. 
Firstly, there is a contribution from $E^{(0)}=U-J$ too.
Secondly, for $f$-states the exact contribution proportional to $E^{(3)}$ is $33/112 \approx 0.295$ instead of $1/4$.
These two corrections can in principle be compensated by using a somewhat larger effective $E^{(3)}$ parameter in OP-B. However thirdly, in OP-B only the $z$ components are included (tensor components $00$).
This is due to that spin mixing are neglected.
This will be found to be a more severe restriction, since in general we can get an independent contribution from each of the three spin components. 
All these three corrections lead to an effectively stronger OP than what was originally suggested by Brooks.
In addition we observe that the OP is essentially two different terms; one
that favours broken time reversal symmetry states with
$\mathbf{w}^{10}\neq0$ (OP-odd), while the second term involving  $\mathbf{w}^{11}$ does not break time reversal symmetry (OP-even). 

There have been other attempts to improve on the OP-B formula \cite{shick2,OP} by starting from an integer occupation in the same spirit as the original work \cite{Brooks}. 
However, since  there are uncertainties in this limit what should be attributed to OP, the resulting formulations usually involve other terms of Eq.~\ref{X-mom} and there 
have never been a general description of OP for a $f$ shell. 
The only approach we are aware of for non-integer systems also failed to get an expression for $f$ systems \cite{solovyev}.

Let us return to the LDA+U calculations of Fig.\ (\ref{fig1:Pu}) where the different energy contributions arising from the different tensor products are given. Here the two angular momenta $kp$ of the double tensor  are coupled into a third $r$, as discussed in Refs.~\cite{Racah-II,vanderLaan}, giving rise to an irreducible tensor $\mathbf{w}^{kpr}$.
As can be seen only a few tensors have any significant contribution to the total exchange energy. They are $\mathbf{w}^{000}$ (total 5$f$ occupation), 
$\mathbf{w}^{011}$ (SP) and 
$\mathbf{w}^{110}$ (OP-even). The OP-odd term $\mathbf{w}^{101}$ is almost detectable for low $U$ values. It is worth noting that the total exchange energy calculated by Eqs.~(\ref{ldau}) and (\ref{X-mom}) are indistinguishable. It is evident from the graph that the OP-even term takes over the SP exchange energy when increasing the effective Coulomb interaction $U$. In fact, as the AMF double counting corresponds to neglecting the contributions to the HF exchange from $\mathbf{w}^{000}$ and 
$\mathbf{w}^{011}$, we observe that the OP-even term solely  determines the HF exchange for all values of $U$! 
{By studying the calculated observables as displayed in bottom part of Fig.\ (\ref{fig1:Pu}) it is worth stressing that although there is a small increase in the $5f$ occupancy from 5.2-5.5, it has little influence, in contrast to what has earlier been assumed \cite{Shick,Anisimov}. It is the steep increase in the magnitude of $\mathbf{w}^{110}$  that stabilises the non-magnetic state.}

The identified broken symmetry of the calculated state is quite intruiging. The order parameter $\mathbf{w}^{110}$ corresponds to that the three components of the spin currents orbit around their different spin quantization axes with equal magnitudes. This leads to a time reversal invariant, scalar, order parameter, and since it arises from spin currents it is a quantity difficult to observe directly in experiments. 

Recently there have, however, been reports on measurements on the branching ratio for the $d$ to $5f$ transition for several actinide systems \cite{vdLaan1,Moore}, from which values of $\mathbf{w}^{110}$ can be obtained through a sum rule \cite{vdLaan2}. These measurements report very large values, not least for $\alpha$-Pu. In the subsequent discussion they attribute this to the strong spin-orbit coupling which brings the 5$f$ states close to a {\em jj} coupling scheme. 
In the light of our finding we would like to alter that analysis slightly. While the spin-orbit coupling is important in the actinides, it is not strong enough to bring the 5$f$ states into a {\em jj}-limit by itself. In fact without the HF term of Eq.\ (\ref{ldau}), i.e.\ in the LDA limit, we calculate a spin-orbit-only value of $-2.4$ while in the presence of the HF term we get enhanced values varying between $-4.4$ and $-7.2$, as seen in Fig.\ (\ref{fig1:Pu}). The values for large $U$ parameter are close to saturation, as indicated in Fig.\ (\ref{fig1:Pu}), which would correspond to $-\frac{4}{3}\mathbf{w}^{000}$. These values should be compared to the measured value of $-5.1$ for Pu in its $\alpha$-phase, assuming a $f^5$ configuration \cite{vdLaan1}. We notice that the exchange term is essential to bring the calculated $\mathbf{w}^{110}$ to the same magnitude as the experimental value. 
This leads to the conclusion that there is a strong competition between {\em different exchange channels} in the actinides, where the spin-orbit coupling plays a role since it favours the OP-even channel over the SP channel. This in accordance with calculations on other actinide systems, where we have found that the $\mathbf{w}^{110}$ always have a large contribution, even for magnetic systems \cite{Cricchio}.
%


As a summary, we conclude that while all calculations essentially involve the dominant SP and OP contributions to exchange,  the main difference between the calculations leading to a magnetic state \cite{Soderlind1,Soderlind2} and the ones leading to a non-magnetic state \cite{Shick,Anisimov}  is their treatments of the OP term, the former utilizes OP-B without any spin-mixing, while the latter uses the correct OP-even including the spin-mixing contribution.
This symmetry broken state, with non-zero spin currents, has a surprisingly simple structure since all non-trivial exchange energy goes  into the OP-even channel. Since the exchange energy usually is larger than the correlation energy it is likely that this state has many similarities with the ground state of $\delta$-Pu.

The derived multi-pole expansion of the exchange energy is general and valid for all types of open shells.
A more general study of the relevant exchange terms for other $f$ and $d$ systems will be published elsewhere \cite{Cricchio}. 
 
This work illustrates a large advantage with the multi-pole expansion of the exchange energy of Eq.\ (\ref{X-mom}) -- it brings forward the physically important exchange channels in a simple way. It is clear that most part of the terms in the expansion has little or no contribution. 
Hence, in the future it is of great interest to perform similar multi-pole expansions on the terms relevant for studies of correlation within the $f$ shell, e.g.\ the Green's function and the self-energy. In particular, in order to better understand $\delta$-Pu and its intriguing electronic structure, one ought to study the effect of correlations on the spin currents of the OP-even state, by means of e.g.\ DMFT calculations together with a similar analysis as performed here.

The support from the Swedish Research Council (VR) is thankfully acknowledged. The computer calculations have been performed at the Swedish high performance centers NSC and UPPMAX under grants provided by the Swedish National Infrastructure for Computing (SNIC).



\begin{thebibliography}{99}
\bibitem{comment} We use the conjugate of the definition used in Ref.\ \cite{vanderLaan}.
\end{thebibliography}
\end{document}